\documentclass[12pt]{article}

\usepackage[left=2cm,right=2cm,top=2cm,bottom=2cm]{geometry}
\usepackage[utf8]{inputenc}
\usepackage{color,soul,xcolor}
\usepackage{amsmath,amssymb}
\usepackage{authblk}
\usepackage{dsfont}
\usepackage{hyperref}
\usepackage{graphicx}

\def\be{\begin{equation}}
\def\ee{\end{equation}}

\numberwithin{equation}{section}

\begin{document}

\title{Perturbative calculation of energy levels for the Dirac equation with generalised momenta}

\author[1]{Marco Maceda\thanks{mmac@xanum.uam.mx}}
\author[1]{Jairo Villafuerte-Lara\thanks{villafuerte@xanum.uam.mx}}
\affil[1]{Departamento de Física, Universidad Autónoma Metropolitana-Iztapalapa \authorcr Av. San Rafael Atlixco 186, C.P. 03340, Deleg. Iztapalapa, Mexico City, México}

\maketitle

\begin{abstract}
We analyse a modified Dirac equation based on a noncommutative structure in phase space. The noncommutative structure induces generalised momenta and contributions to the energy levels of the standard Dirac equation. Using techniques of perturbation theory, we use this approach to find the lowest order corrections to the energy levels and eigenfunctions for two linear potentials in three dimensions, one with radial dependence and another with a triangular shape along one spatial dimension. We find that the corrections due to the noncommutative contributions may be of the same order as the relativistic ones.
\end{abstract}

\section{Introduction}
\label{intro}

In recent years, most of the theories that attempt to describe the universe at the quantum level had experienced significant developments; among them, we find string theory, quantum loops, spin foams and theories based on noncommutative geometry. All of them give us insight on the issue of how to unify general relativity with quantum theory. 

On the other hand, grand unification theories have remarkable features and to be able to distinguish the main characteristics of each of these theories, we have to take into account the phenomenology of quantum gravity. For this purpose, we may use certain types of astrophysical phenomena in which we can detect individual variations in their energy or in their observation time that are natural consequences of the theory.  Indeed, there are indications that this type of situations may be present in phenomena such as the GZK cut-off where variations of the relationship between energy and momenta exist~\cite{Sidharth:2015qaf}, or a generalised uncertainty principle comes into play~\cite{Ghosh:2012cv}. This class of phenomena will also help to infer if specific features of some theory of quantum gravity are having a non-trivial influence at low energies as pointed out in~\cite{AmelinoCamelia:1997gz}.

The introduction of a minimum length scale is a structural feature within some theories of quantum gravity~\cite{AmelinoCamelia:2000ge,Maggiore:1993kv,Girelli:2004md}. The existence of a minimum length is intimately related to generalised uncertainty principles~\cite{Ghosh:2012cv,Ali:2009zq} or as a consequence of a modified energy-momentum dispersion relation~\cite{Freidel:2005me,Sidharth:2015qaf}. The concept of generalised momenta incorporates into a theory, a minimum length in a straightforward way and hence, quantum gravity effects up to a certain extent. 

For these reasons, we consider a modified Dirac equation along the lines of~\cite{Ghosh:2012cv}. Since the Dirac equation is an essential pillar within particle physics and solid-state physics, it is interesting to analyse a modification to this equation using the formalism of generalised momenta. Generalised momenta satisfy commutation relations in phase space that are analogous to those first introduced by Snyder~\cite{Snyder:1946qz}; they are seen as more fundamental, reducing to the standard momenta in a specific limit. In our work, we consider that the commutators involving coordinate and momenta operators depend quadratically on the momenta; this specific dependence is the result of a noncommutative theory where a minimum length exists. Then, generalised momenta incorporate quantum gravity like-effects in a semiclassical way. For convenience and also for computational purposes, we consider potentials that depend linearly in one coordinate variable. More specifically, we work with a linear radial potential in spherical coordinates and with a triangular linear potential along the $z$-axis in Cartesian coordinates. Both potentials are discussed in detail in the literature mainly because of their relevance in particle physics, as illustrated in confinement models of quarkonium~\cite{Lichtenberg:1987ms}, and also in connection with solid-state physics where this class of potential is used to model the interaction of an electric field inside a metal~\cite{Castro:2010yb,Ashcroft:2011}. Furthermore, these potentials are suitable for a perturbative approach of energy levels based on the exact solutions of the associated Schrödinger equation~\cite{Rutkowski:1986}; the use of the large and small components of the Dirac equation, after some algebraic manipulations, leads to closed expressions for the perturbative corrections.

We use generalised momenta in our approach as an attempt to find quantum gravity like-corrections that may be susceptible to observation at low energies.  We should mention that modified Dirac equations that incorporate the breaking of Lorentz and CPT invariances also exist in the literature; when applied to quantum systems like the hydrogen atom~\cite{Kostelecky:1999zh,Lehnert:2004ri,Ferreira:2006kg,Kharlanov:2007yp,Yoder:2012ks}, they lead to modified dispersion relations and modified energy levels.

Our work is structured as follows. In Sec.~\ref{secc:2}, we review the implications of a Generalised Uncertainty Principle (GUP) and its relation with a minimum length scale. We also briefly review in Sec.~\ref{secc:3} the perturbation theory for the Dirac equation and we extend it in Sec.~\ref{secc:4}  to the case of generalised momenta, finding the first-order corrections to the energy levels. In Sec.~\ref{secc:5}, we apply this approach to the specific cases of radial and triangular linear potentials. In the Conclusions section, we comment on these results and give some perspectives on future work.

\section{ Generalised phase space and minimal length scale}
\label{secc:2}

Generalised phase space operators are a consequence of GUTs and quantum gravity effects where it is natural to consider generalised momenta; one of their main consequences is a modified dispersion relation where the relativistic energy get corrections involving quartic or higher powers of the momentum. The structure for a generalised momentum operator is~\cite{AmelinoCamelia:1997gz} 
\be
\mathbf{P}_{i}=p_{i}\left(1-a_{0}\frac{p}{E_{p}/c}+b_{0}\frac{p^{2}}{\left(E_{p}/c\right)^{2}}\right), \qquad i=1, 2, 3,
\ee
where $a_{0},b_{0}$ are arbitrary constants, $E_{p}$ is Planck's energy, $p_i$ is the standard $i$-th momentum operator and $p := \sqrt{\sum_{i}p_{i}^{2}}$; we now define
\be
a := \frac{a_{0}}{E_{p}/c}, \qquad b := \frac{b_{0}}{\left(E_{p}/c\right)^{2}}.
\ee
The constants $a, b$ are suppressed by Planck's energy and therefore, we assume them to be small in the following; the generalized momentum operator is then
\be
\mathbf{P}_{i}=p_{i}\left(1-ap+bp^{2}\right).
\label{genp}
\ee
Accordingly, we also define a generalised coordinate operator as
\be
\mathbf{X}_i := x_i,  \qquad i=1, 2, 3,
\label{genx}
\ee
where the coordinate operators $x_i$ satisfy the standard commutation relations
\be
\left[x_i , p_j \right]=i\hbar\delta_{ij}.
\ee
The generalised operators $(\mathbf{X}_i, \mathbf{P}_j)$ are seen as more fundamental than the operators $(x_i, p_j)$; they reduce to the latter when $a, b \to 0$. It is straightforward to see that these generalised operators satisfy the following commutation relations
\be
\left[\mathbf{X}_{i}, \mathbf{P}_{j}\right] = i\hbar\left[ \delta_{ij} - a\left( \delta_{ij}p+\frac{p_{i}p_{j}}{p}\right) + b\left(\delta_{ij}p^{2}+2p_{i}p_{j} \right) \right]. 
\label{crgm}
\ee
The right hand side of these commutators becomes a function of the generalised coordinate and momentum operators once we invert Eqs.~(\ref{genp}) and~(\ref{genx}). A direct and interesting consequence of the above relation is the existence of a minimal length scale $\ell_{min}$; to see how this length scale arises consider the one dimensional reduction of Eq.~(\ref{crgm}), namely 
\be
\left[\mathbf{X} , \mathbf{P} \right] = i\hbar\left( 1 - 2a p + 3b p^2 \right).
\label{crgm1d}
\ee
From the definition of the generalised momentum $\mathbf P$, we obtain
\be
p = \mathbf{P} \left[ 1+ a\mathbf{P} +(2a^2 - b)\mathbf{P}^{2} \right],
\ee
to linear terms in $b$ and quadratic terms on $a$. Therefore, up to these orders, Eq.~(\ref{crgm1d}) becomes
\be
\left[\mathbf{X} , \mathbf{P} \right] = i\hbar\left[ 1 - 2a \mathbf{P}+ (3b - 2a^2) \mathbf{P}^2 \right].
\label{gcr}
\ee
Using now the Heisenberg-Robertson uncertainty relation~\cite{Heisenberg:1927zz,Robertson:1929zz}
\be
\Delta \mathbf{X} \Delta \mathbf{P} \geq \left| \frac{1}{2} \left\langle \left[ \mathbf{X}, \mathbf{P} \right] \right\rangle \right|,
\ee
we obtain
\be
\Delta \mathbf{X} \Delta \mathbf{P} \geq \frac{\hbar}{2}\left(1 - 2a\left\langle \mathbf{P} \right\rangle + \left(3b - 2a^{2}\right)\left\langle \mathbf{P}^{2}\right\rangle \right),
\label{gup}
\ee
where the bracket indicates an expectation value of the corresponding operator. If we set $b=2a^{2}$, we obtain the Generalised Uncertainty Principle (GUP) proposed in~\cite{Ali:2009zq}.

To find the minimal length scale, we first replace $\left\langle \mathbf{P} \right\rangle \rightarrow\Delta \mathbf{P}$ and $\left\langle \mathbf{P}^{2}\right\rangle \rightarrow\left(\Delta \mathbf{P}\right)^{2}$ in Eq.~(\ref{gup}); therefore
\be
\Delta \mathbf{X} \Delta \mathbf{P} \geq \frac{\hbar}{2} \left[ 1-2a\Delta \mathbf{P} + \left(3b-2a^{2}\right)\left(\Delta \mathbf{P} \right)^{2} \right].
\ee
This GUP is saturated when 
\be
\Delta \mathbf{X} = \frac{\hbar}{2} \left[ \frac{1}{\Delta \mathbf{P}} - 2a + \left(3b-2a^{2}\right) \Delta \mathbf{P}  \right],
\ee
and for a minimum for $\Delta \mathbf{X}$ to exist, we need that
\be
\frac{\partial\Delta\mathbf{X} }{\partial\Delta \mathbf{P}} \Big|_{(\Delta \mathbf{P})_{ext}} = \frac{\hbar}{2} \left[ -\frac{1}{(\Delta \mathbf{P})^2_{ext} } + 3b-2a^{2} \right] = 0.
\ee
Assuming $3b-2a^{2} > 0$, we obtain the value $(\Delta \mathbf{P})_{ext} = \left(3b-2a^{2}\right)^{-1/2}$; in consequence the minimal length scale is 
\be
\ell_{min} := (\Delta\mathbf{X}) _{ext} = \hbar ( \sqrt{3b-2a^{2}} - a).
\ee
The scale length is always positive provided that $b \geq a^2$.

We finally mention that using the Schrödinger representation for the operators $x_i$ and $p_i$, we obtain a representation for the generalised operators $\mathbf X_i$ and $\mathbf P_i$; when writing the Dirac equation with generalised momenta, this representation may be helpful. 

\section{Perturbation theory for the Dirac equation}
\label{secc:3}

In this section we fix notation and also review the perturbation theory for the Dirac equation~\cite{Rutkowski:1986} that we shall use later. We consider first the Dirac equation with a scalar potential $V$ in the form
\begin{equation}
\left( \mathrm{i}\hbar c\gamma^{\mu}\frac{\partial}{\partial x^{\mu}} - (mc^{2} + V) \right)\psi =0,
\end{equation}
where
\be
\gamma^0 =\left( \begin{array}{cc}
\mathds{1}_{2\times 2} & 0
\\[4pt]
0 & -\mathds{1}_{2\times 2}
\end{array} \right)
\qquad
\gamma^k =\left( \begin{array}{cc}
0 & \sigma_{k}
\\[4pt]
-\sigma_{k} & 0
\end{array} \right), \qquad k = 1, 2, 3,
\ee
with $\mathds{1}_{2\times 2}$ the unit matrix of dimension two, $\sigma_k$ the Pauli sigma matrices and $x^{\mu}=(ct,\vec{x} )$; $\psi$ is the relativistic wave function. Multiplication by $\beta := \gamma^0$ and the definition $\alpha^i := \gamma^0 \gamma^i$ lead to
\begin{equation}
\left[ \mathrm{i}\hbar \frac{\partial}{\partial t}+\mathrm{i}\hbar c\alpha^{i}\frac{\partial}{\partial x^{i}} - \beta \left( mc^{2}+ V\right) \right]\psi=0,
\end{equation}
or equivalently
\begin{equation}
\mathrm{i}\hbar \frac{\partial}{\partial t}\psi = -\mathrm{i}\hbar c\alpha^{i}\frac{\partial}{\partial x^{i}} + \beta \left( mc^{2}+ V\right)\psi.
\end{equation}
Defining now the momenta $p_i := - \mathrm{i}\hbar \displaystyle\frac {\partial}{\partial x^i}$, we have the Dirac equation
\begin{equation}
\mathrm{i}\hbar \frac{\partial}{\partial t}\psi =\left[ c \boldsymbol{\alpha} \cdot \boldsymbol{p} + \beta \left( mc^{2}+ V\right) \right]\psi,
\end{equation}
with a scalar potential $V$~\cite{Rafelski:1973fn,Alhaidari:2005fw}. For time-independent potentials, this equation becomes
\begin{equation}
\left[ c \boldsymbol{\alpha} \cdot \boldsymbol{p} + \beta \left( mc^{2}+ V\right)- E \right]\psi=0.
\end{equation}
We now introduce hermitian projectors $\beta^\pm$ according to
\begin{equation}
\beta^{\pm}: = \frac{\mathds{1}_{4\times 4} \pm \beta}{2}, \qquad \mathds{1}_{4\times 4}=\beta^{+}+\beta^{-}, \qquad \beta^\pm \beta = \pm \beta^\pm,
\end{equation}
and we rewrite the Dirac equation as
\begin{equation}
\left[H_{0}-\beta^{+} E_{n}+  \beta^{-}(- V - E_{n})\right]\psi=0, \qquad H_{0} := c \boldsymbol{\alpha} \cdot \boldsymbol{p} - 2mc^{2}\beta^{-} +\beta^{+} V,
\label{diracequ}
\end{equation}
where $E_n := E - mc^2$ and we redefined $\beta V \to V$. This form of the Dirac equation is suitable to perform perturbation theory as pointed out in~\cite{Rutkowski:1986}, where $V$ should not necessarily be small but rather we only need to identify a suitable constant associated to our physical system as the perturbation parameter. For this purpose, the unperturbed system is defined by the equation
\begin{equation}
\left( H_{0} - \beta^{+} E_{n}^{(0)} \right)\psi_{n}=0.
\end{equation}
If we make the decomposition $\psi_n := (\phi_{n}^{(0)+}, \phi_{n}^{(0)-})^T$ where $\phi_{n}^{(0)\pm}$ are the large and small two-component spinors of $\psi_n$, we have
\begin{equation}
\left(\begin{array}{cc}
V \mathds{1}_{2\times 2}& c \boldsymbol{\sigma} \cdot \boldsymbol{p}
\\[4pt]
c \boldsymbol{\sigma} \cdot \boldsymbol{p} & -2mc^{2}
\end{array} \right) \left( \begin{array}{c}
\phi_{n}^{(0)+}
\\
\phi_{n}^{(0)-}
\end{array} \right)=\left(\begin{array}{cc}
E_{n}^{(0)} & 0
\\[4pt]
0 & 0
\end{array} \right) \left( \begin{array}{c}
\phi_{n}^{(0)+}
\\[4pt]
\phi_{n}^{(0)-}
\end{array} \right).
\end{equation}
We obtain then a system of coupled equations for the large and small components
\begin{equation}
V \phi_{n}^{(0)+} + c (\boldsymbol{\sigma} \cdot \boldsymbol{p})\phi_{n}^{(0)-}=E_{n}^{(0)} \phi_{n}^{(0)+}, 
\qquad
 c( \boldsymbol{\sigma} \cdot \boldsymbol{p})\phi_{n}^{(0)+}-2mc^{2}\phi_{n}^{(0)-}=0,
\end{equation}
that is equivalent to the relations
\begin{equation}
\left(  \frac{p^{2}}{2m} + V \right)\phi_{n}^{(0)+}=E_{n}^{(0)} \phi_{n}^{(0)+},
\qquad
\phi_{n}^{(0)-}= \frac{(\boldsymbol{\sigma} \cdot \boldsymbol{p})}{2mc}\phi_{n}^{(0)+}.
\label{bigsmallcomp}
\end{equation}
The large component $\phi_{n}^{(0)+}$ satisfy the time-independent Schrödinger equation with potential $V$; from it we can calculate the small component $\phi_{n}^{(0)-}$ straightforwardly. 

If we write now
\begin{equation}
\left[H_{0}-\beta^{+} E_{n}+ \lambda \beta^{-}(-V - E_{n}) \right]\psi_{n}=0,
\end{equation}
where $\lambda$ is a perturbation parameter, and substitute of the series development 
\begin{eqnarray}
\psi_{n} &=& \psi_{n}^{(0)}+\lambda \psi_{n}^{(1)}+\lambda^{2}\psi_{n}^{(2)}+...,
\nonumber \\[4pt]
E_{n} &=& E_{n}^{(0)}+\lambda E_{n}^{(1)}+\lambda^{2}E_{n}^{(2)}+...,
\label{series}
\end{eqnarray}
in the previous expression, we arrive to a set of coupled equations for the perturbations; to zero order on $\lambda$ the relevant equation is
 \begin{equation}
 \left( H_{0} - \beta^{+}E_{n}^{(0)} \right)\psi_{n}^{(0)}=0,
 \end{equation}
and to first order 
\begin{equation}
\left( H_{0} - \beta^{+}E_{n}^{(0)} \right)\psi_{n}^{(1)}+ \left[ -\beta^{+} E_{n}^{(1)}+\beta^{-}(-V - E_{n}^{(0)}) \right] \psi_{n}^{(0)}=0.
\end{equation} 
By choosing the normalization of the unperturbed wave function $\psi_{n}^{(0)}$ as
\begin{equation}
\langle \psi_{n}^{(0)+},\psi_{n}^{(0)+}\rangle=\langle \psi_{n}^{(0)},\beta^{+}\psi_{n}^{(0)}\rangle=1,
\end{equation} 
we obtain the correction to the eigen-energies $E_n^{(0)}$ to first order on $\lambda$ as~\cite{Rutkowski:1986} 
\begin{equation}
E_{n}^{(1)} = \langle \psi_{n}^{(0)}, \beta^{-}(-V-E_{n}^{(0)}) \psi_{n}^{(0)}\rangle = -\langle \psi_{n}^{(0)-}, (V + E_{n}^{(0)}) \psi_{n}^{(0)-}\rangle,
\end{equation}
where the second equality follows from the hermitian character of $\beta^-$.

\section{The Dirac equation with generalised momenta and perturbation theory}
\label{secc:4}

We now extend the results in the previous section when using generalised momenta; we choose the latter as
\begin{equation}
\mathbf{P}_i=p_i (1-a^{2}p^{2}), \qquad i = 1, 2 ,3.
\end{equation}
Here $a$ is an arbitrary small parameter; in the context of quantum gravity, we may identify $a^{2} = \left( \frac{l_{P}}{\hbar} \right)^{2}$ where $l_P$ is Planck's length.  a minimal length. Notice that no quadratic dependence on the standard momentum operator $p$ is present; as we illustrate below, this choice has the advantage of allowing explicit calculations for the noncommutative contributions. According to the discussion in Sec.~\ref{secc:2}, we have a noncommutative structure in phase space and a minimal length scale in our model if $a^2 \neq 0$. 

We write the time-independent Dirac equation with generalised momenta as
\begin{equation}
\left[ c \boldsymbol{\alpha} \cdot \mathbf{P} + \beta \left( mc^{2}+ V\right) - E_n \right]\psi=0,
\label{diracequmod}
\end{equation}
or in terms of the standard momenta $p_i$
\begin{equation}
\left[ c \boldsymbol{\alpha} \cdot \mathbf{p} + \beta \left( mc^{2}+ V\right) - a^2 c (\boldsymbol{\alpha} \cdot \mathbf p) p^2 - E_n \right]\psi=0.
\label{diracequmod2}
\end{equation}
In this form, we have a modified Dirac equation where the third term inside the brackets provides corrections from a noncommutative structure of phase space due to a quantum theory of gravity for instance. 

To apply the perturbative approach of the previous section to Eq.~(\ref{diracequmod2}), we rewrite it using the projectors $\beta^\pm$ as
\begin{equation}
\left[H_{0}-\beta^{+} E_{n}+ \beta^{-}(-V - E_{n}) - a^2 c (\boldsymbol{\alpha} \cdot \mathbf p) p^2 \right]\psi_{n}=0,
\end{equation}
where $H_0$ is defined in Eq.~(\ref{diracequ}). The terms involving $\beta^-$ and $a^2$ are then perturbations; accordingly we write
\begin{equation}
\left[H_{0}-\beta^{+} E_{n}+ \lambda \{ \beta^{-}(-V - E_{n}) - a^2 c (\boldsymbol{\alpha} \cdot \mathbf p) p^2 \} \right]\psi_{n}=0,
\end{equation}
and consider a series development as in Eq.~(\ref{series}). It follows that the first order corrections to the eigen-energies are
\begin{eqnarray}
E_{n}^{(1)} &=& \langle \psi_{n}^{(0)},(\beta^{-}(-V - E_{n}^{(0)})-a^{2} c (\boldsymbol{\alpha} \cdot \boldsymbol{p})p^{2})\psi_{n}^{(0)}\rangle
\nonumber \\[4pt]
&=& -\langle \psi_{n}^{(0)-},(V + E_{n}^{(0)})\psi_{n}^{(0)-}\rangle -a^{2} c\langle \psi_{n}^{(0)}, (\boldsymbol{\alpha} \cdot \boldsymbol{p})p^{2} \psi_{n}^{(0)}\rangle.
\label{eemod}
\end{eqnarray}
At first sight, we need to evaluate some derivatives of the wave function to determine this quantity completely, in particular for the second contribution in Eq.~(\ref{eemod}). We now proceed to this calculation; from the Schrödinger equation satisfied by $\phi_{n}^{(0)+}$, we have
\begin{equation}
\frac{p^{2}}{2m} \phi_{n}^{(0)+}=(E_{n}^{(0)}-V)\phi_{n}^{(0)+},
\end{equation}
and recalling the relation between large and small components in Eq.~(\ref{bigsmallcomp}), we calculate then
\begin{eqnarray}
(\boldsymbol{\alpha} \cdot \boldsymbol{p})p^{2} \psi_{n}^{(0)}\rangle &=& \left( \begin{array}{cc}
0 & \boldsymbol{\sigma} \cdot \boldsymbol{p}
\\
\boldsymbol{\sigma} \cdot \boldsymbol{p} & 0
\end{array} \right) p^{2}  \left( \begin{array}{c}
\phi_{n}^{(0)+}
\\
\phi_{n}^{(0)-}
\end{array}\right)
\nonumber \\[4pt]
&=&
\left(\begin{array}{c}
(\boldsymbol{\sigma} \cdot \boldsymbol{p}) p^{2} \phi_{n}^{(0)-}
\\
(\boldsymbol{\sigma} \cdot \boldsymbol{p}) p^{2} \phi_{n}^{(0)+}
\end{array} \right)
=
\left(\begin{array}{c}
\displaystyle\frac{1}{2mc}(\boldsymbol{\sigma} \cdot \boldsymbol{p}) p^{2} (\boldsymbol{\sigma} \cdot \boldsymbol{p}) \phi_{n}^{(0)+}
\\
(\boldsymbol{\sigma} \cdot \boldsymbol{p}) 2m (E^{(0)}_{n}-V) \phi_{n}^{(0)+}
\end{array} \right).
\end{eqnarray}
Interchanging $p^{2}$ with $(\boldsymbol{\sigma} \cdot \boldsymbol{p})$ in the last expression leads to
\begin{eqnarray}
(\boldsymbol{\alpha} \cdot \boldsymbol{p})p^{2} \psi_{n}^{(0)}\rangle &=& \left(\begin{array}{c}
\displaystyle\frac{1}{2mc} p^{2} (\boldsymbol{\sigma} \cdot \boldsymbol{p}) (\boldsymbol{\sigma} \cdot \boldsymbol{p}) \phi_{n}^{(0)+}
\\[4pt]
(\boldsymbol{\sigma} \cdot \boldsymbol{p}) 2m (E^{(0)}_{n}-V) \phi_{n}^{(0)+}
\end{array} \right)
=
\left(\begin{array}{c}
\displaystyle\frac{1}{2mc} p^{2} p^{2} \phi_{n}^{(0)+}
\\
(\boldsymbol{\sigma} \cdot \boldsymbol{p}) 2m (E^{(0)}_{n}-V) \phi_{n}^{(0)+}
\end{array} \right)
\nonumber \\[4pt]
&=&
\left(\begin{array}{c}
\displaystyle\frac{1}{c} p^{2} (E^{(0)}_{n}-V) \phi_{n}^{(0)+}
\\
(\boldsymbol{\sigma} \cdot \boldsymbol{p}) 2m (E^{(0)}_{n}-V) \phi_{n}^{(0)+}
\end{array} \right),
\end{eqnarray}
where we used $(\boldsymbol{\sigma} \cdot \boldsymbol{p})(\boldsymbol{\sigma} \cdot \boldsymbol{p})= p^{2}$ together with the Schrödinger equation satisfied by the large component $\phi_{n}^{(0)+}$. With this result, we obtain for the expectation value
\begin{eqnarray}
\langle \psi_{n}^{(0)}, (\boldsymbol{\alpha} \cdot \boldsymbol{p}) p^{2} \psi_{n}^{(0)}\rangle &=&
\langle ( \phi_{n}^{(0)+},\phi_{n}^{(0)-}), \left(\begin{array}{c}
\displaystyle\frac{1}{c} p^{2} (E^{(0)}_{n}-V) \phi_{n}^{(0)+}
\\
(\boldsymbol{\sigma} \cdot \boldsymbol{p}) 2m (E^{(0)}_{n}-V) \phi_{n}^{(0)+}
\end{array} \right)\rangle
\nonumber \\[4pt]
&=&\langle \phi_{n}^{(0)+}, \frac{1}{c} p^{2} (E^{(0)}_{n}-V) \phi_{n}^{(0)+}  \rangle
+\langle \phi_{n}^{(0)-}, (\boldsymbol{\sigma} \cdot \boldsymbol{p}) 2m (E^{(0)}_{n}-V) \phi_{n}^{(0)+}  \rangle. 
\end{eqnarray}
Since $p^{2}$ and $(\boldsymbol{\sigma} \cdot \boldsymbol{p})$ are hermitian operators, we can simplify the above expression as follows
\begin{eqnarray}
\langle \psi_{n}^{(0)}, (\boldsymbol{\alpha} \cdot \boldsymbol{p})p^{2} \psi_{n}^{(0)}\rangle &=& \langle \frac{p^{2}}{c}  \phi_{n}^{(0)+},  (E^{(0)}_{n}-V) \phi_{n}^{(0)+}  \rangle
+\langle (\boldsymbol{\sigma} \cdot \boldsymbol{p}) \phi_{n}^{(0)-},  2m (E^{(0)}_{n}-V) \phi_{n}^{(0)+}  \rangle
\nonumber \\[4pt]
&=& \frac{2m}{c}\langle (E^{(0)}_{n}-V)  \phi_{n}^{(0)+},  (E^{(0)}_{n}-V) \phi_{n}^{(0)+}  \rangle
+ \frac{1}{2mc}\langle p^{2} \phi_{n}^{(0)+},  2m (E^{(0)}_{n}-V) \phi_{n}^{(0)+}  \rangle
\nonumber \\[4pt]
&=& \frac{2m}{c}\langle  \phi_{n}^{(0)+},  (E^{(0)}_{n}-V)^{2} \phi_{n}^{(0)+}  \rangle
+ \frac{2m}{c}\langle  (E^{(0)}_{n}-V) \phi_{n}^{(0)+},  (E^{(0)}_{n}-V) \phi_{n}^{(0)+}  \rangle
\nonumber \\[4pt]
&=& \frac{2m}{c}\langle  \phi_{n}^{(0)+},  (E^{(0)}_{n}-V)^{2} \phi_{n}^{(0)+}  \rangle
+ \frac{2m}{c}\langle  \phi_{n}^{(0)+},  (E^{(0)}_{n}-V)^{2} \phi_{n}^{(0)+}  \rangle
\nonumber \\[4pt]
&=& \frac{4m}{c}\langle  \phi_{n}^{(0)+},  (E^{(0)}_{n}-V)^{2} \phi_{n}^{(0)+}  \rangle.
\end{eqnarray}
In consequence, the perturbation $E_n^{(1)}$ has the final form
\be
E_n^{(1)} = -\langle \phi_{n}^{(0)-},(V + E_{n}^{(0)})\phi_{n}^{(0)-}\rangle -4 m a^{2} \langle  \phi_{n}^{(0)+},  (E^{(0)}_{n}-V)^{2} \phi_{n}^{(0)+}  \rangle.
\ee
This explicit result for the perturbation on the energy shows that the evaluation of derivatives of the wave function or the potential $V$ is not required for this quantity; the presence of only a linear and cubic dependence of the generalised momenta on the standard momenta is relevant here.

\section{Applications}
\label{secc:5}

Having determined the first order corrections to the energy eigenvalues for the Dirac equation in the presence of generalised momenta, we now turn to the application and explicit calculation of the perturbative corrections for the specific case of scalar linear potentials. As mentioned in the Introduction, these potentials are of significance within the context of quark confinement, solid state and condensed matter~\cite{Lucha:1991vn,Richard:2012xw,Vega:2018eiq}. We introduce these potential in the Dirac equation as scalar potentials instead of vector potential~\cite{Rafelski:1973fn,Tezuka:2013,Alhaidari:2005fw}.

\subsection{Modified Dirac equation with linear radial potential}
\label{secc:5a}

As a first example, we recall that in spherical coordinates $(r, \theta, \phi)$ the Schrödinger equation 
\be
\left[ \frac {p^2}2 + V(r) \right] \Psi = E \Psi,
\ee
with a linear radial potential $V(r) = qr$ becomes the equation
\be
\left\{ \frac 12 \left[ - \frac {d^2}{dr^2} + \frac {l(l+1)}{r^2} \right] + qr \right\} \phi= E \phi,
\ee
with the substitution $\Psi (r, \theta, \phi) = \phi(r) Y_{lm} (\theta, \phi)/r$. Analytic solutions exist for $l=0$ and numerical solutions for other values; in the former case, we have~\cite{Moshinsky:2002na}
\be
\phi_n (r) = Ai (z), \qquad z := (2q)^{1/3} r + z_n,
\ee
where $z_n < 0$ is the $n$-th zero of the Airy function $Ai(z)$. Furthermore, the energy eigenvalues are
\be
E_n = - \frac {(2q)^{2/3}}2 z_n = - \frac {q^{2/3}}{2^{1/3}} z_n,
\ee
and they are always positive. The normalised wave function is then
\be
\Psi_n (r, \theta, \phi) = \alpha \frac {(2q)^{1/6}}{\sqrt{4\pi}} \frac {Ai[(2q)^{1/3} r + z_n]}r = \alpha \sqrt{ \frac {2q}{4\pi}} \frac {Ai(z)}{z - z_n}.
\ee
where $\alpha := \left( \int_{z_n}^\infty Ai(\xi)^2 d\xi \right)^{-1/2} = [Ai^\prime(z_n)]^{-1}$.

Consider now the Dirac equation with generalised momenta in the presence of the linear radial potential. According to the perturbation theory developed in Sec.~\ref{secc:4}, we write $\phi_n^{(0)+} = (\Psi_n , 0)^T$ for the large component and $E_n^{(0)} = - q^{2/3} z_n /2^{1/3}$ for the unperturbed energy eigenvalues. With these identifications, we calculate first
\begin{eqnarray}
\langle \phi_n^{(0)+} , (E_n^{(0)} - V)^2 \phi_n^{(0)+} \rangle &=& \int_0^\infty \int_0^\pi \int_0^{2\pi} \, \Psi_n(r, \theta, \phi)^2  (E_n^{(0)} - qr)^2 r^2 \sin \theta dr d\theta d\phi,
\nonumber \\[4pt]
&=& \alpha^2 \frac {q^{4/3}}{2^{2/3}} \int_{z_n}^\infty z^2 Ai(z)^2 dz = \alpha^2 \frac {q^{4/3}}{2^{2/3}} \times \frac 15 [Ai^\prime(z_n)]^2 z_n^2,
\nonumber \\[4pt]
&=& \frac 15 \frac {q^{4/3}}{2^{2/3}} z_n^2,
\end{eqnarray}
where we used 
\be
E_n^{(0)} - q r = - \frac {q^{2/3}}{2^{1/3}} z_n - q \frac {z - z_n}{(2q)^{1/3}} = - \frac {q^{2/3}}{2^{1/3}} z.
\ee
We now proceed to the calculation of the small component $\phi_n^{(0)-}$; using spherical coordinates, we have
\begin{eqnarray}
\phi_n^{(0)-} &=& \frac 1{2c} (\boldsymbol\sigma \cdot \boldsymbol p) \phi_n^{(0)+} = -\frac i{2c} \sigma^r (\partial_r \Psi_n, 0)^T 
\nonumber \\[4pt]
&=& - \frac i{2c} \frac {\alpha (2q)^{1/6}}{\sqrt{4\pi}}\sigma^r \left( \frac {d}{dr} \frac {Ai[(2q)^{1/3} r + z_n]}r, 0 \right)^T
\nonumber \\[4pt]
&=& - \frac i{2c} \frac {\alpha (2q)^{5/6}}{\sqrt{4\pi}}\sigma^r \left( \frac {d}{dz} \frac {Ai(z)}{z - z_n}, 0 \right)^T,
\end{eqnarray}
where $\sigma^r = \left( \begin{array}{cc}\cos \theta & \sin\theta e^{-i\phi} \\[4pt] \sin\theta e^{i\phi} & -\cos\theta \end{array} \right)$~\cite{Wang:2014}; notice that $\sigma^{r \, \dagger} = \sigma^r$ and $(\sigma^r)^2 = \mathds{1}_{2\times 2}$. In consequence we obtain
\begin{eqnarray}
\langle \phi_n^{(0)-} , (E_n^{(0)} + V) \phi_n^{(0)-} \rangle &=& \int_0^\infty \int_0^\pi \int_0^{2\pi} \, \frac 1{(2c)^2} (\partial_r \Psi_n, 0) \sigma^{r \, \dagger} (E_n^{(0)} + q r ) \sigma^r (\partial_r \Psi_n, 0)^T r^2 \sin \theta dr d\theta d\phi
\nonumber \\[4pt]
&=& \alpha^2 \frac {q^{4/3}}{2^{2/3}c^2} \int_{z_n}^\infty \left( \frac z2 - z_n \right) (z - z_n)^2 \left( \frac {d}{dz} \frac {Ai(z)}{z - z_n} \right)^2 dz
\nonumber \\[4pt]
&=&\alpha^2 \frac {q^{4/3}}{2^{2/3}c^2} \times \left( -\frac 56 [Ai^\prime(z_n)]^2 z_n \right) = -\frac 56 \frac {q^{4/3}}{2^{2/3}c^2} z_n,
\end{eqnarray}
where we used
\be
E_n^{(0)} + q r = - \frac {q^{2/3}}{2^{1/3}} z_n + q \frac {z - z_n}{(2q)^{1/3}} = \frac {q^{2/3}}{2^{1/3}} z - (2q)^{2/3} z_n = (2q)^{2/3} \left( \frac z2 - z_n \right).
\ee
The first order correction to the energy eigenvalues is then
\be
E_n^{(1)} = \frac 56 \frac {q^{4/3}}{2^{2/3}c^2} z_n - 4a^2 \times \frac 15 \frac {q^{4/3}}{2^{2/3}} z_n^2 = \frac {q^{4/3}}{2^{2/3}} \left( \frac 1{c^2} \frac 56 z_n - a^2 \frac 45 z^2_n \right).
\ee
Since $z_n$ is always negative, we have that this correction corresponds to a total energy eigenvalue smaller than that of the unperturbed case. Furthermore, the contribution due to the generalised momenta is always negative; if $a \sim c^{-1}$, then this contribution is of the same order as the relativistic one for small values of $n$ but becomes an order higher as $n$ increases.

\begin{figure}[htbp]
\begin{center}
\includegraphics{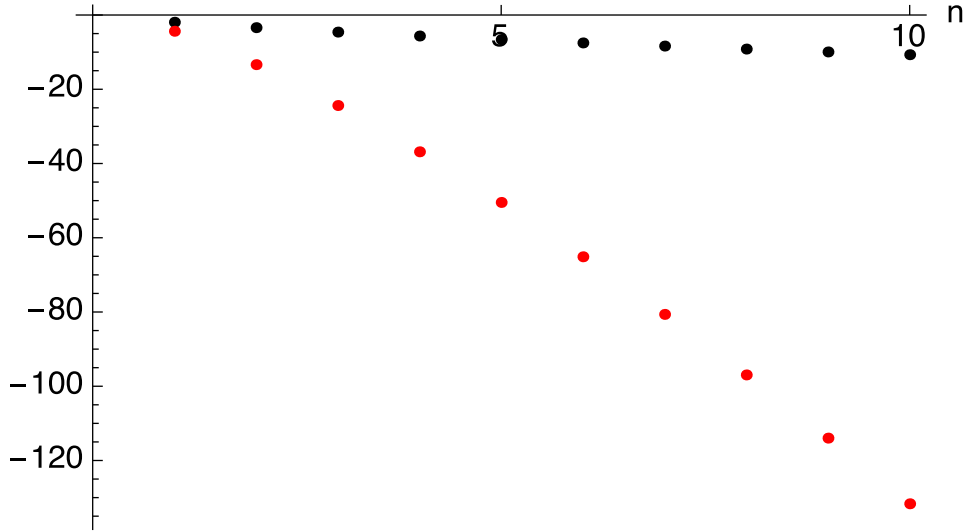}
\caption{Behaviour of $5 z_n/6$ (black points) and $-4z_n^2/5$ (red points) as a function of $n$.}
\label{default}
\end{center}
\end{figure}

\subsection{Modified Dirac equation with triangular potential}
\label{secc:5b}

As a more complex problem, we consider the case of a triangular potential. Recall first that the Schrödinger equation with one-dimensional triangular potential along the $z$-axis is~\cite{Castro:2010yb}
\begin{equation}
V(z)= \frac{V_{0}}{L} \left( \vert z \vert - L \right) \left[ \Theta(z+L)- \Theta(z-L) \right], 
\label{triangularpot}
\end{equation}
where $\Theta(w)$ is the Heaviside function, is solvable in the real line; due to its invariance under $z\to -z$, it suffices to discuss the solutions in $\mathbb R^+$ only. For the interval $0 < z < L$ the wave function is
\be
\psi(\zeta) = c_a Ai(\zeta)+ c_{b}Bi(\zeta),
\ee
where $Ai(w), Bi(w)$ are Airy's functions of the first and second kind; in the previous expression we have
\be
\zeta := \frac{v_{0}^{1/3}}{L}\left[ \vert z \vert - L\left(1 + \frac{\varepsilon}{v_{0}} \right)  \right], 
\ee
with $\epsilon := 2m L^2 E/\hbar^2, v_0 := 2mL^2 V_0/\hbar^2$. 

There are even/odd parity wave functions defined respectively by the constraints~\cite{Castro:2010yb}
\be
c_{a}Ai'(\zeta_{0})+c_{b}Bi'(\zeta_{0})=0, \qquad c_{a}Ai(\zeta_{0})+c_{b}Bi(\zeta_{0})=0,
\ee
where $\zeta_{0}$ is the value of $\zeta$ at $x=0$. On the other hand, when $L < z$, the wave function is
\be
\psi(z)=c \exp \left(- \frac{\sqrt{-\varepsilon}}{L}z \right).
\ee
The matching conditions for $\psi$ and its derivative at $z = L$, or equivalently at $\zeta_L = \zeta_0 + v_0^{1/3}$, allow us to determine the eigen-energies as~\cite{Castro:2010yb}
\begin{equation}
E= -V_{0}\left[ 1+ \zeta_{0}\left( \frac{\hbar^{2}}{2mL^{2}V_{0}} \right)^{1/3} \right].
\end{equation}
The value $\zeta_0$ in the previous expression is found by solving the transcendental equation
\begin{equation}
\frac{Ai'(\zeta_{L})-\alpha Ai(\zeta_{L})}{Bi'(\zeta_{L})-\alpha Bi(\zeta_{L})}= \begin{cases} Ai'(\zeta_{0})/Bi'(\zeta_{0}), 
\\[4pt]
Ai(\zeta_{0})/Bi(\zeta_{0}), \qquad \alpha := -(\zeta_0 + v_0^{1/3})^{1/2},
\end{cases}
\end{equation}
for even/odd parities respectively. Furthermore, we can also determine the relation among the different constants $c_a, c_b$ and $c$; for even parity we have
\be
c_b = - c_a Ai'(\zeta_{0})/Bi'(\zeta_{0}), \qquad c = c_a e^{\sqrt{|\epsilon|}} \left[ Ai(\zeta_L) - Bi(\zeta_L) Ai'(\zeta_{0})/Bi'(\zeta_{0}) \right],
\ee
while for odd parity
\be
c_b = - c_a Ai(\zeta_{0})/Bi(\zeta_{0}), \qquad c = c_a e^{\sqrt{|\epsilon|}} \left[ Ai(\zeta_L) - Bi(\zeta_L) Ai(\zeta_{0})/Bi(\zeta_{0}) \right].
\ee
The constant $c_a$ is then fixed by the normalisation condition $\displaystyle\int_{-\infty}^\infty |\psi(z) |^2 \,dz = 1$.

Consider now the three-dimensional situation where the scalar potential in the three-dimensional Dirac equation is 
\be
V(x, y, z) = V_1(x) V_2(y) V(z), 
\ee
where $V_1(x), V_2(y)$ are infinite potentials of width $L_x$ and $L_y$ respectively, and $V(z)$ is the same as in Eq.~(\ref{triangularpot}). We write then the large component of the relativistic wave function to zero-order as 
\be
\phi_n^{(0)+} (\vec x) = \frac 2{\sqrt{L_x L_y}} \sin(p_x x) \sin (p_y y) \varphi_n^{(0)+} (z) \, \boldsymbol \xi, 
\ee
where $\boldsymbol \xi$ is a bi-spinor\footnote{For even parity solutions we take $\boldsymbol \xi = (1, 0)^T$ and for odd parity solutions $\boldsymbol \xi = (0, 1)^T$.}, $\varphi_n^{(0)+}(z)$ encodes the dependence on $z$, and $p_x := n \pi/L_x, p_y := m\pi/L_y, n, m \in \mathbb Z$. It follows that the equation satisfied by $\varphi_n^{(0)+}(z)$ is
\be
\left[ \frac 1{2m} p_z^2 + V(z) \right] \varphi_n^{(0)+} = E_{n\parallel}^{(0)} \varphi_n^{(0)+}, \qquad E_{n\parallel}^{(0)} := E_n^{(0)} - E_{n\perp}^{(0)},  \qquad E_{n\perp}^{(0)} := \frac 1{2m} (p_x^2 + p_y^2).
\ee
Therefore, we have as solution
\begin{eqnarray}
\varphi_{n, even/odd}^{(0)+}(z) &=& \Theta(+z) \left\{ \Theta(L - z)[c_a Ai(\zeta)+ c_{b}Bi(\zeta)] + \Theta(z - L) c \exp (-\sqrt{|\epsilon_{n\parallel}|}z/L) \right\}
\nonumber \\[4pt]
&&\pm \Theta(-z) \left\{ \Theta(L + z)[c_a Ai(\zeta)+ c_{b}Bi(\zeta)] + \Theta(-L - z) c \exp (\sqrt{|\epsilon_{n\parallel}|}z/L) \right\},
\end{eqnarray}
where $\epsilon_{n\parallel} := 2m L^2 E_{n\parallel}/\hbar^2$. On the other hand, in terms of $E_\parallel$ and $E_\perp$, the first order correction becomes
\begin{eqnarray}
E_n^{(1)} &=& -\langle \phi_{n}^{(0)-},(E_{n\parallel}^{(0)} + V) \phi_{n}^{(0)-} \rangle - 4 m a^{2} \langle \phi_{n}^{(0)+},  (E^{(0)}_{n\parallel}-V)^{2} \phi_{n}^{(0)+} \rangle 
\nonumber \\[4pt]
&&- 8 m a^{2} E_\perp \langle \phi_{n}^{(0)+},  (E^{(0)}_{n\parallel}-V) \phi_{n}^{(0)+} \rangle - E_\perp \langle \phi_{n}^{(0)-}, \phi_{n}^{(0)-}\rangle - 4 m a^{2} E_\perp^2.
\end{eqnarray}

If we define $v(z):= 2mL^2 V(z)/\hbar^2$, fix $\hbar = m = L =1$, and consider the specific case when the transverse momentum $\vec p_\perp := (p_x, p_y)$ vanishes, then we may write
\begin{eqnarray}
\langle \varphi_{n}^{(0)-},(E_{n\parallel}^{(0)} + V) \varphi_{n}^{(0)-}\rangle &=& \frac 1{4c^2} \int (\epsilon_\parallel^{(0)+} + v) (\partial_z \varphi_n^{(0)+})^2 =: \frac 1{4c^2} \langle  \epsilon_\parallel^{(0)} + v \rangle_{--},
\nonumber \\[4pt]
\langle  \phi_{n}^{(0)+},  (E^{(0)}_{n\parallel}-V)^{2} \phi_{n}^{(0)+}  \rangle &=& \frac 12 \int (\epsilon_\parallel^{(0)} - v)^2 (\varphi_n^{(0)+})^2 =: \frac 12 \langle  (\epsilon_\parallel^{(0)} - v)^2 \rangle_{++},
\nonumber \\[4pt]
\langle  \phi_{n}^{(0)+},  (E^{(0)}_{n\parallel}-V) \phi_{n}^{(0)+}  \rangle &=& \int (\epsilon_\parallel^{(0)} - v) (\varphi_n^{(0)+})^2 =: \langle  \epsilon_\parallel^{(0)} - v \rangle_{++}.
\end{eqnarray}
Accordingly, we have
\be
E_n^{(1)} = -\frac 1{4c^2} \langle  \epsilon_\parallel^{(0)} + v \rangle_{--} - 2 a^{2} \langle  (\epsilon_\parallel^{(0)} - v)^2 \rangle_{++}. 
\ee
In Table~\ref{data} we give the numerical values for the first order corrections to the ground states associated to even parity wave functions. We notice from them that the magnitude of $\langle  \epsilon_\parallel^{(0)} + v \rangle_{--}$ increases in absolute value as $v_0$ gets bigger, nevertheless this contribution is suppressed by a relativistic factor; we remark that the quantity $\langle  (\epsilon_\parallel^{(0)} - v)^2 \rangle_{++}$ also shows an increase but it is less pronounced. If $a \sim c^{-1}$ then both contributions are of the same order of magnitude but they have opposite signs, therefore counterbalancing with each other; this behaviour may be susceptible of experimental observation. 

\section{Conclusions}

Using the notion of generalised momenta, we analysed a modified Dirac equation in the presence of linear scalar potentials. For this purpose, we applied a suitable form of perturbation theory; a parameter $a$ encoded the higher than quadratic dependence on the momentum introduced by the generalised momenta. The corrections, relativistic and generalised, involved the small and large components of the relativistic wave function.

To exemplify our approach, we discussed the case of linear potentials.  In a first instance, we analysed a radial potential in three dimensions in the specific situation of zero angular momentum; this potential is useful in the context of quark confinement and analytic expressions for the standard Dirac wave functions exist. Using them, we applied the perturbation theory developed in Sec. 4 and determined the perturbed eigenvalues of the energy levels. We found explicit expressions for these eigenvalues: the relativistic correction depends linearly on the zeroes of the Airy function Ai(z), but the generalised momentum correction depends quadratically on them; they combine to give an overall negative correction to the zeroth-order values of the energy levels.

As a second application, we analysed in three dimensions a linear triangular potential with dependence on one coordinate and with an infinite well potential in the two others. Even though this problem seemed to be more accessible at first sight, the calculation of the energy levels associated with the contributions of the generalised momenta required numerical calculations. For this example, the overall correction to the energy levels may be positive or negative depending on the ratio $c/a$.

The approach developed here may be useful in finding constraints on theories based on generalised momenta; its extension to a general curved spacetime background and other systems such as ultra-cold neutrons is worthwhile pursuing. By looking into experiments performed at low energies, we may set up bounds to the deformation parameters and find hints of interactions happening at higher energy levels.

\section*{Acknowledgments}
J. Villafuerte-Lara acknowledges support from Universidad Autónoma Metropolitana (UAM).

\begin{table}[htp]
\caption{Triangular potential along the $z$-axis: Numerical data for ground states}
\centering
\begin{tabular}{|c|c|c|c|c|c|c|c|c|}
\hline \hline
$v_0$ & $z_0$ & $\epsilon_\parallel^{(0)}$ & $c_a$ & $c_b$ & $c$ & $\langle  \epsilon_\parallel^{(0)} - v \rangle_{++}$ & $\langle (\epsilon_\parallel^{(0)} - v)^2 \rangle_{++}$ & $\langle  \epsilon_\parallel^{(0)} + v \rangle_{--}$ \\[4pt]
\hline
1&-0.826181&-0.173819&1.02392&0.170583&0.661594&0.123379&0.132641&-0.0364646 \\[4pt]
2&-0.918052&-0.542683&1.30148&0.116604&0.918542&0.3108&0.578932&-0.290765\\[4pt]
3&-0.956877&-1.00962&1.46048&0.0813997&1.12559&0.494445&1.24057&-0.869706\\[4pt]
4&-0.977389&-1.53713&1.56976&0.0588883&1.3129&0.665659&2.04381&-1.79791\\[4pt]
5&-0.98958&-2.10645&1.65245&0.0439097&1.4908&0.824469&2.94726&-3.07349\\[4pt]
6&-0.997388&-2.7067&1.71884&0.033551&1.66409&0.972408&3.92678&-4.68648\\[4pt]
7&-1.00266&-3.33096&1.77433&0.0261487&1.83537&1.1111&4.96745&-6.62507\\[4pt]
8&-1.00636&-3.97456&1.82204&0.0207187&2.00625&1.24194&6.05954&-8.87749\\[4pt]
9&-1.00904&-4.63414&1.86397&0.0166413&2.17775&1.3661&7.19637&-11.4331\\[4pt]
10&-1.01103&-5.30723&1.90142&0.0135208&2.35058&1.48452&8.37319&-14.2822\\[4pt]
\hline
\end{tabular}
\label{data}
\end{table}

\newpage


\begin{thebibliography}{10}

\bibitem{Sidharth:2015qaf}
B.~G. Sidharth, A.~Das, and A.~D. Roy, ``{Lorentz Invariance Violation:
  Modification of the Compton Scattering and the GZK Cutoff and Other
  Effects},''
\href{http://dx.doi.org/10.1007/s10773-015-2889-3}{{\em Int. J. Theor. Phys.}
  {\bfseries 55} no.~5, (2016) 2541--2547}.

\bibitem{Ghosh:2012cv}
S.~Ghosh, ``{Generalized Uncertainty Principle, Modified Dispersion Relation
  and Barrier Penetration by a Dirac Particle},''
  \href{http://dx.doi.org/10.1007/s10773-014-2265-8}{{\em Int. J. Theor. Phys.}
  {\bfseries 54} no.~3, (2015) 736--748},
\href{http://arxiv.org/abs/1202.1962}{{\ttfamily arXiv:1202.1962 [hep-th]}}.

\bibitem{AmelinoCamelia:1997gz}
G.~Amelino-Camelia, J.~R. Ellis, N.~E. Mavromatos, D.~V. Nanopoulos, and
  S.~Sarkar, ``{Tests of quantum gravity from observations of gamma-ray
  bursts},'' \href{http://dx.doi.org/10.1038/31647}{{\em Nature} {\bfseries
  393} (1998) 763--765},
\href{http://arxiv.org/abs/astro-ph/9712103}{{\ttfamily arXiv:astro-ph/9712103
  [astro-ph]}}.

\bibitem{AmelinoCamelia:2000ge}
G.~Amelino-Camelia, ``{Testable scenario for relativity with minimum length},''
  \href{http://dx.doi.org/10.1016/S0370-2693(01)00506-8}{{\em Phys. Lett.}
  {\bfseries B510} (2001) 255--263},
\href{http://arxiv.org/abs/hep-th/0012238}{{\ttfamily arXiv:hep-th/0012238
  [hep-th]}}.

\bibitem{Maggiore:1993kv}
M.~Maggiore, ``{The Algebraic structure of the generalized uncertainty
  principle},'' \href{http://dx.doi.org/10.1016/0370-2693(93)90785-G}{{\em
  Phys. Lett.} {\bfseries B319} (1993) 83--86},
\href{http://arxiv.org/abs/hep-th/9309034}{{\ttfamily arXiv:hep-th/9309034
  [hep-th]}}.

\bibitem{Girelli:2004md}
F.~Girelli, E.~R. Livine, and D.~Oriti, ``{Deformed special relativity as an
  effective flat limit of quantum gravity},''
  \href{http://dx.doi.org/10.1016/j.nuclphysb.2004.11.026}{{\em Nucl. Phys.}
  {\bfseries B708} (2005) 411--433},
\href{http://arxiv.org/abs/gr-qc/0406100}{{\ttfamily arXiv:gr-qc/0406100
  [gr-qc]}}.

\bibitem{Ali:2009zq}
A.~F. Ali, S.~Das, and E.~C. Vagenas, ``{Discreteness of Space from the
  Generalized Uncertainty Principle},''
  \href{http://dx.doi.org/10.1016/j.physletb.2009.06.061}{{\em Phys. Lett.}
  {\bfseries B678} (2009) 497--499},
\href{http://arxiv.org/abs/0906.5396}{{\ttfamily arXiv:0906.5396 [hep-th]}}.

\bibitem{Freidel:2005me}
L.~Freidel and E.~R. Livine, ``{3D Quantum Gravity and Effective Noncommutative
  Quantum Field Theory},''
  \href{http://dx.doi.org/10.1103/PhysRevLett.96.221301}{{\em Phys. Rev. Lett.}
  {\bfseries 96} (2006) 221301},
  \href{http://arxiv.org/abs/hep-th/0512113}{{\ttfamily arXiv:hep-th/0512113
  [hep-th]}}.
[Bulg. J. Phys.33,no.s1,111(2006)].

\bibitem{Snyder:1946qz}
H.~S. Snyder, ``{Quantized space-time},''
\href{http://dx.doi.org/10.1103/PhysRev.71.38}{{\em Phys. Rev.} {\bfseries 71}
  (1947) 38--41}.

\bibitem{Lichtenberg:1987ms}
D.~B. Lichtenberg, ``{Energy Levels of Quarkonia in Potential Models},''
\href{http://dx.doi.org/10.1142/S0217751X87000879}{{\em Int. J. Mod. Phys.}
  {\bfseries A2} (1987) 1669}.

\bibitem{Castro:2010yb}
L.~B. Castro and A.~S. de~Castro, ``{On the bound-state spectrum of a
  nonrelativistic particle in the background of a short-ranged linear
  potential},'' {\em Electron. J. Theor. Phys.} {\bfseries 7} no.~23, (2010)
  155--160,
\href{http://arxiv.org/abs/1003.2993}{{\ttfamily arXiv:1003.2993 [quant-ph]}}.

\bibitem{Ashcroft:2011}
N.~Ashcroft and N.~Mermin, {\em Solid State Physics}.
\newblock Cengage Learning, 2011.
\newblock \url{https://books.google.com.mx/books?id=x\_s\_YAAACAAJ}.

\bibitem{Rutkowski:1986}
A.~Rutkowski, ``Relativistic perturbation theory. i. a new perturbation
  approach to the dirac equation,''
  \href{http://dx.doi.org/10.1088/0022-3700/19/2/005}{{\em Journal of Physics
  B: Atomic and Molecular Physics} {\bfseries 19} no.~2, (Jan, 1986) 149--158}.
  \url{https://doi.org/10.1088%2F0022-3700%2F19%2F2%2F005}.

\bibitem{Kostelecky:1999zh}
V.~A. Kostelecky and C.~D. Lane, ``{Nonrelativistic quantum Hamiltonian for
  Lorentz violation},'' \href{http://dx.doi.org/10.1063/1.533090}{{\em J. Math.
  Phys.} {\bfseries 40} (1999) 6245--6253},
\href{http://arxiv.org/abs/hep-ph/9909542}{{\ttfamily arXiv:hep-ph/9909542
  [hep-ph]}}.

\bibitem{Lehnert:2004ri}
R.~Lehnert, ``{Dirac theory within the standard model extension},''
  \href{http://dx.doi.org/10.1063/1.1769105}{{\em J. Math. Phys.} {\bfseries
  45} (2004) 3399--3412},
\href{http://arxiv.org/abs/hep-ph/0401084}{{\ttfamily arXiv:hep-ph/0401084
  [hep-ph]}}.

\bibitem{Ferreira:2006kg}
M.~M. Ferreira, Jr. and F.~M.~O. Moucherek, ``{Influence of Lorentz- and
  CPT-violating terms on the Dirac equation},''
  \href{http://dx.doi.org/10.1142/S0217751X06033842}{{\em Int. J. Mod. Phys.}
  {\bfseries A21} (2006) 6211--6227},
\href{http://arxiv.org/abs/hep-th/0601018}{{\ttfamily arXiv:hep-th/0601018
  [hep-th]}}.

\bibitem{Kharlanov:2007yp}
O.~G. Kharlanov and V.~C. Zhukovsky, ``{CPT and Lorentz violation effects in
  hydrogen-like atoms},'' \href{http://dx.doi.org/10.1063/1.2785123}{{\em J.
  Math. Phys.} {\bfseries 48} (2007) 092302},
\href{http://arxiv.org/abs/0705.3306}{{\ttfamily arXiv:0705.3306 [hep-th]}}.

\bibitem{Yoder:2012ks}
T.~J. Yoder and G.~S. Adkins, ``{Higher order corrections to the hydrogen
  spectrum from the Standard-Model Extension},''
  \href{http://dx.doi.org/10.1103/PhysRevD.86.116005}{{\em Phys. Rev.}
  {\bfseries D86} (2012) 116005},
\href{http://arxiv.org/abs/1211.3018}{{\ttfamily arXiv:1211.3018 [hep-ph]}}.

\bibitem{Heisenberg:1927zz}
W.~a~Heisenberg, ``{Uber den anschaulichen Inhalt der quantentheoretischen
  Kinematik und Mechanik},''
\href{http://dx.doi.org/10.1007/BF01397280}{{\em Z. Phys.} {\bfseries 43}
  (1927) 172--198}.

\bibitem{Robertson:1929zz}
H.~P. Robertson, ``{The Uncertainty Principle},''
\href{http://dx.doi.org/10.1103/PhysRev.34.163}{{\em Phys. Rev.} {\bfseries 34}
  (1929) 163--164}.

\bibitem{Rafelski:1973fn}
J.~Rafelski, G.~Soff, B.~Muller, and W.~Greiner, ``{Solution of the Dirac
  Equation for Scalar Potentials and its Implications in Atomic Physics},''
{\em Z. Naturforsch.} {\bfseries 28A} (1973) 1389.

\bibitem{Alhaidari:2005fw}
A.~D. Alhaidari, H.~Bahlouli, and A.~Al-Hasan, ``{The Dirac and Klein-Gordon
  equations with equal scalar and vector potentials},''
  \href{http://dx.doi.org/10.1016/j.physleta.2005.09.008}{{\em Phys. Lett.}
  {\bfseries A349} (2006) 87--97},
\href{http://arxiv.org/abs/hep-th/0503208}{{\ttfamily arXiv:hep-th/0503208
  [hep-th]}}.

\bibitem{Lucha:1991vn}
W.~Lucha, F.~F. Schoberl, and D.~Gromes, ``{Bound states of quarks},''
\href{http://dx.doi.org/10.1016/0370-1573(91)90001-3}{{\em Phys. Rept.}
  {\bfseries 200} (1991) 127--240}.

\bibitem{Richard:2012xw}
J.-M. Richard, ``{An introduction to the quark model},'' in {\em {Ferrara
  International School Niccolò Cabeo 2012: Hadronic spectroscopy Ferrara,
  Italy, May 21-26, 2012}}.
\newblock 2012.
\newblock
\href{http://arxiv.org/abs/1205.4326}{{\ttfamily arXiv:1205.4326 [hep-ph]}}.
\newblock

\bibitem{Vega:2018eiq}
A.~Vega and F.~Rojas, ``{Confinement potentials for the study of heavy
  mesons},''
\href{http://arxiv.org/abs/1810.08080}{{\ttfamily arXiv:1810.08080 [hep-ph]}}.

\bibitem{Tezuka:2013}
H.~Tezuka, ``Analytical solutions of the dirac equation with a scalar linear
  potential,'' \href{http://dx.doi.org/10.1063/1.4820388}{{\em AIP Advances}
  {\bfseries 3} no.~8, (2013) 082135},
  \href{http://arxiv.org/abs/https://doi.org/10.1063/1.4820388}{{\ttfamily
  https://doi.org/10.1063/1.4820388}}. \url{https://doi.org/10.1063/1.4820388}.

\bibitem{Moshinsky:2002na}
M.~Moshinsky and A.~Suarez~Moreno, ``{The spectra of a Hamiltonian with a
  linear radial potential derived by a variational calculation based on a set
  of harmonic oscillator states},''
{\em Rev. Mex. Fis.} {\bfseries 48} (2002) 39--42.

\bibitem{Wang:2014}
Y.-L. Wang, L.~Du, C.-T. Xu, X.-J. Liu, and H.-S. Zong, ``Pauli equation for a
  charged spin particle on a curved surface in an electric and magnetic
  field,'' \href{http://dx.doi.org/10.1103/PhysRevA.90.042117}{{\em Phys. Rev.
  A} {\bfseries 90} (Oct, 2014) 042117}.
  \url{https://link.aps.org/doi/10.1103/PhysRevA.90.042117}.

\end{thebibliography}

\providecommand{\href}[2]{#2}\begingroup\raggedright\endgroup

\end{document}